\documentclass[aps,prd,preprint,superscriptaddress]{revtex4-2}

\usepackage{ulem}
\usepackage{xcolor}
\usepackage[colorlinks=true]{hyperref}
\usepackage{soul}
\usepackage{booktabs}
\usepackage{mathrsfs}
\usepackage{amssymb}

\usepackage{amsmath,amssymb,color,epsfig,graphicx}
\allowdisplaybreaks[4]

\newcommand{\be}{\begin{equation}}
\newcommand{\ee}{\end{equation}}
\newcommand{\bea}{\setlength\arraycolsep{2pt} \begin{eqnarray}}
\newcommand{\eea}{\end{eqnarray}}
\newcommand{\nn}{\nonumber}

\def\0{{\sst{(0)}}}
\def\1{{\sst{(1)}}}
\def\2{{\sst{(2)}}}
\def\3{{\sst{(3)}}}
\def\4{{\sst{(4)}}}
\def\5{{\sst{(5)}}}
\def\6{{\sst{(6)}}}
\def\7{{\sst{(7)}}}
\def\8{{\sst{(8)}}}
\def\sst#1{{\scriptscriptstyle #1}}

\begin{document}

\hypersetup{
    linkcolor=blue,
    citecolor=red,
    urlcolor=magenta
}


\title{Geodesics and shadows of the spindle-deformed Kerr black hole}



\author{Hong-Da Lyu}
\email[]{hongdalyu@sdu.edu.cn}
\affiliation{Key Laboratory of Particle Physics and Particle Irradiation (MOE),
Institute of Frontier and Interdisciplinary Science,
Shandong University, Qingdao, Shandong, 266237, China}

\author{Mingzhi Wang}
\email[]{wmz9085@126.com}
\affiliation{School of Mathematics and Physics, Qingdao University of Science and Technology,
Qingdao, Shandong 266061, Peoples Republic of China}

\author{Shoulong Li}
\email[Corresponding author: ]{shoulongli@hunnu.edu.cn}
\affiliation{Department of Physics, Key Laboratory of Low Dimensional Quantum Structures and Quantum Control of Ministry of Education, and Institute of Interdisciplinary Studies, Hunan Normal University, Changsha, 410081, China}
\affiliation{Hunan Research Center of the Basic Discipline for Quantum Effects and Quantum Technologies, Hunan Normal University, Changsha, 410081, China}


\date{\today}

\begin{abstract}

Recently, a new exact Ricci-flat rotating black-hole solution was constructed in four-dimensional general relativity, in which an additional parameter $B$ characterizes a spindle deformation of the Kerr geometry. We study geodesic motion and black-hole shadows in this spacetime. The Hamilton-Jacobi equation is not exactly separable for either timelike or null geodesics. Remarkably, however, at the leading nontrivial order, ${\cal O}(B^2)$, null but not timelike geodesics become separable. In the timelike sector, the spindle deformation shifts the innermost stable circular orbit and can give rise to an outermost stable circular orbit. In the null sector, exploiting the perturbatively separated equations, we analytically determine the photon region, equatorial photon orbits, and black-hole shadow, and compare the resulting predictions with direct ray tracing in the exact spacetime. The numerical results validate the perturbative treatment and quantify the deviations from Kerr.

\end{abstract}


\maketitle


\section{Introduction}

The Kerr spacetime~\cite{Kerr:1963ud} occupies a distinguished position in general relativity (GR), serving both as the standard reference model for interpreting astrophysical observations of rotating black holes and as a theoretical laboratory for understanding strong-field gravity and the spacetime structure of black holes. Within four-dimensional GR, nontrivial Ricci-flat extensions of the Kerr geometry are rather limited, with important examples including Kerr-NUT (Newman-Unti-Tamburino) and accelerating Kerr geometries, which introduce additional parameters and enrich our understanding of black-hole spacetimes~\cite{Stephani:2003tm,Griffiths:2009dfa}. Recently, Ma and L\"u~\cite{Ma:2026otg} (ML) constructed a new exact Ricci-flat rotating black-hole solution in which an additional parameter $B$ introduces a nontrivial deformation of the Kerr geometry, with the Kerr solution recovered when $B$ vanishes.

The appearance of the parameter $B$ is closely tied to the construction of this class of Ricci-flat black-hole solutions. It first emerged in the static $B$-deformed Schwarzschild black hole obtained by Astorino~\cite{Astorino:2026okd}, through an Ernst transformation applied to a Schwarzschild black hole immersed in an external Bertotti-Robinson electromagnetic field~\cite{Astorino:2025lih,Ernst:1976mzr}. Although $B$ originally characterizes the strength of the background electromagnetic field in the corresponding Einstein-Maxwell solution, it survives in the final Ricci-flat metric after the field is removed, becoming a residual geometric parameter that deforms the Schwarzschild geometry and renders the spacetime non-asymptotically flat. The rotating ML solution is constructed in a similar spirit, through a demagnetizing transformation~\cite{Ernst:1976mzr,Ernst:1976bsr,Gibbons:2013yq} applied to the recently reported Kerr-Bertotti-Robinson (KBR) black hole~\cite{Podolsky:2025tle,Ovcharenko:2025cpm}, with the additional subtlety that the closed timelike curves induced by rotation must be appropriately treated. In the final ML black hole, $B$ is likewise no longer an external electromagnetic-field strength, but an integration parameter characterizing an intrinsic spindle deformation of the Kerr geometry.

This new class of exact black-hole solutions has motivated several recent studies of the role of the parameter $B$. In particular, their thermodynamics has been investigated, and a consistent first law was shown to hold after an appropriate normalization of the Killing frame~\cite{Ma:2026otg,Astorino:2026okd}. This analysis not only identifies the physical mass and angular momentum relevant for subsequent studies of the black-hole geometry, but also sharpens the distinction between the Ricci-flat ML black hole and its KBR seed~\cite{Hu:2026slp}: in the former, $B$ is not associated with an independent matter charge, but enters the physical quantities through the non-asymptotically flat structure of the spacetime. Beyond thermodynamics, the $B$-deformed black hole has also motivated further discussions on solution-generating transformations and the interpretation of the resulting spacetimes~\cite{Herdeiro:2026jem,Ma:2026ima,Ovcharenko:2026uxi,Ma:2026uok}. Weak-field phenomenological constraints on the static spindle-deformed geometry have also been investigated~\cite{Yu:2026tqk}.

Despite these developments, how the spindle deformation affects the geodesic structure around a rotating black hole remains much less understood. Geodesic motion provides a direct link between spacetime geometry and observable strong-field phenomena~\cite{Abramowicz:2011xu,Perlick:2021aok,Cunha:2018acu,Chen:2022scf,Wang:2022kvg,EventHorizonTelescope:2019dse,EventHorizonTelescope:2022wkp,Johannsen:2015hib}: timelike geodesics determine the orbital dynamics and stability of matter in accretion flows, with the innermost stable circular orbit (ISCO) setting a characteristic inner scale of thin accretion disks, while null geodesics govern photon propagation, strong-field lensing, photon rings, and black-hole shadows. While some aspects of timelike geodesics have been investigated in the static case~\cite{Astorino:2026okd}, the issue becomes more subtle for rotating black holes, because the geodesic structure of a rotating spacetime is closely tied to the separability of the Hamilton-Jacobi equation and the associated hidden symmetry~\cite{Chandrasekhar:1985kt,Carter:1968rr}. In the Kerr spacetime, the Hamilton-Jacobi equation is separable for both massive and massless particles. For comparison, in the KBR seed geometry of the rotating ML black hole, separability has been shown only for null geodesics~\cite{Wang:2025vsx} (see also Refs.~\cite{Zeng:2025tji,Wang:2025bjf,Zhang:2025ole,Gray:2025lwy,Andersson:2025bhq,Xamidov:2026kqs,Wan:2026lca,Li:2026zxk} for further studies of geodesic motion in the KBR background). It is therefore natural to ask whether the demagnetized Ricci-flat ML geometry inherits this separability property or departs from it, and whether the spindle deformation gives rise to qualitatively new geodesic behavior.

In this work, we study geodesic motion and black-hole shadows in the spindle-deformed Kerr black hole, namely the ML black hole. We show that the Hamilton-Jacobi equation is not exactly separable for either massive or massless particles. This differs from Kerr, where both sectors are separable, and from the KBR seed, where null separability is retained. Remarkably, however, null separability is recovered at the leading nontrivial order, ${\cal O}(B^2)$. We then investigate how the spindle deformation modifies timelike and null geodesics and the resulting black-hole shadow. The remainder of this paper is organized as follows. In Sec.~\ref{MLsection}, we briefly review the spindle-deformed Kerr spacetime and the corresponding physical mass and angular momentum. In Sec.~\ref{geodesicsection}, we examine the separability of the Hamilton–Jacobi equation and then analyze timelike and null geodesics separately. In Sec.~\ref{shadowsection}, we study the black-hole shadow. Finally, Sec.~\ref{conclusion} summarizes our conclusions.

\section{Spindle-deformed Kerr black holes}\label{MLsection}

In this section, we briefly review the spindle-deformed Kerr solution, namely the ML black hole~\cite{Ma:2026otg}, and introduce the physical mass and angular momentum used in the following analysis. We start from the original form of the metric,
\begin{align}\label{eq:metric}
ds^2=&-\frac{Q}{\mathbb{L}\Omega^4}\left(\frac{a x^2}{P_0}d\phi+ \frac{a^2x^2\mathbb{M}+ \mathbb{L}r^2\Omega^2(1-x^2)P}{r^4(1-x^2)P-a^2x^4Q}dt\right)^2\nonumber\\
&+\frac{(1-x^2)P}{\Omega^4\mathbb{L}}\left(\frac{r^2}{P_0}d\phi+ \frac{\left(r^2\mathbb{M}+ \mathbb{L}Q\Omega^2x^2\right)a}{r^4(1-x^2)P-a^2x^4Q}dt\right)^2 \nn \\
&+\mathbb{L}\left(\frac{dr^2}{Q}+\frac{dx^2}{(1-x^2)P}\right)\,,
\end{align}
where $x=\cos\theta$. The metric functions are
\begin{widetext}
\begin{align}\label{eq:MLfunctions}
\begin{split}
\mathbb{M}=&\frac{1}{2I_1^3\Omega}
\bigg\{
\Omega
\left[
I_1I_2m r x^2
\left(
B^2m r(B^2r^2+x^2)
+
2I_1(B^2r^2+1)
\right)
-
I_1^3\Sigma(B^2r^2x^2+1)
\right]\\
&
+
(B^2r^2+1)
\Big[
I_1^3\Sigma(a^2B^2x^2-1)
+
I_2m r x^2
\big(
I_2B^4m^2r^2x^2
+
3I_1I_2B^2m r x^2\\
&-
I_1^2(3a^2B^2x^2+B^2r^2x^2-2)
\big)
\Big]
\bigg\} \,,\\
\mathbb{L}
={}&
\frac{1}{4I_1^2\Omega^4}
\bigg\{
\Sigma
\left[
B^4r^2m^2 I_2 x^2
+
I_1^2
\left(
2+a^2B^4r^2x^2
+
B^2(r^2-a^2x^2-r^2x^2)
\right)
\right]\\
&
+
B^2rm I_2x^2
\left(
2a^2I_1x^2
+
B^2r^3m I_2x^2
+
2r^2I_1(2-a^2B^2x^2)
\right)\\
&
+
2\Omega I_1
\left(
\Sigma I_1
+
B^2r^3m I_2x^2
\right)
\bigg\}  \,,
\end{split}\nn\\
\Sigma =& r^2 + a^2 x^2,
\quad
P = 1 + B^2\left(m^2\frac{I_2}{I_1^2}-a^2\right)x^2,
\quad
Q = (1+B^2r^2)\Delta,
\nonumber\\
\Omega^2 =& (1+B^2r^2)-B^2x^2\Delta,
\quad
\Delta = \left(1-B^2m^2\frac{I_2}{I_1^2}\right)r^2
-2m\frac{I_2}{I_1}r
+a^2 \,.
\end{align}
\end{widetext}
The remaining constants are
\bea
 I_1 &=& 1-\frac{1}{2}B^2a^2 \,,
\quad 
I_2 = 1-B^2a^2 \,, \nn\\ 
P_0 &=& \frac{2I_2(I_1^2+B^2m^2)}{I_1^2(1+\sqrt{I_2})}  \,. \label{eq:P0I1I2}
\eea
Here $P_0$ is fixed by requiring the azimuthal coordinate $\phi$ to have period $2\pi$. The solution is characterized by three integration constants, $m$, $a$, and $B$: the first two are related to the mass and rotation of the black hole, while $B$ is a new intrinsic parameter that controls the spindle deformation of the spacetime. For $a=0$, the solution reduces to the static spindle-deformed black hole obtained in Ref.~\cite{Astorino:2026okd}. For $a=m=0$, the geometry becomes a warped product of $U(1)$ and ${\rm AdS}_3$.  With the above angular identification, however, the metric contains naked closed timelike curves, and this pathology persists even in the limit $B\to0$.  To recover the regular Kerr limit and remove these closed timelike curves, it is necessary to perform the linear coordinate transformation
\be
t=t' -\frac{2a}{(1+\sqrt{I_2})P_0}\phi \,. \label{eq:removectc}
\ee
This linear transformation does not change the horizon locations. Since $g^{rr}=Q/\mathbb{L}$ and $Q=(1+B^2r^2)\Delta$, the horizons are determined by the roots of $\Delta(r)=0$. The inner and outer horizons, $r_-$ and $r_+$, are therefore given by
\be
r_\pm= \frac{\left(1+I_2\right) \left( 2 I_2 m \pm\sqrt{4 I_2 \left(m ^2-a^2\right) -a^6 B^4}\right)}{a^4 B^4+4 I_2 \left(1-B^2 m ^2\right)} \,.
\label{eq:horizons}
\ee
The extremal limit is obtained when the two horizons coincide, namely when
\be
m
=
\frac{ I_1 a}
{\sqrt{I_2}} \,.
\label{extreme}
\ee
This expression reduces to the Kerr extremal condition $m=a$ in the limit $B\to0$.

For the subsequent analysis, it is useful to parameterize the solution by the physical mass and angular momentum rather than by the integration constants $m$ and $a$. This requires a choice of normalization for the Killing coordinates. In asymptotically flat Kerr spacetime, the timelike Killing vector is canonically normalized at infinity. In the present case, the asymptotic geometry is the $B$-spindle rather than Minkowski spacetime, and no such canonical normalization is available. Consequently, the conserved mass and angular momentum depend on the choice of Killing frame. Following Ref.~\cite{Ma:2026otg}, we fix this freedom by demanding that the conserved charges obey the first law of black-hole thermodynamics. This is implemented by the further linear transformation
\be
t' = \lambda_1 t'' \,,\quad \phi=\phi''+\lambda_2 B t'' \,.
\ee
where $\lambda_1$ and $\lambda_2$ are dimensionless constants. Dropping the double primes on $t''$ and $\phi''$ for notational simplicity, the metric becomes
\begin{widetext}
\begin{align}\label{eq:metric3}
ds^2=&-\frac{Q}{\mathbb{L}\Omega^4}\left\{\frac{a x^2}{P_0}(d\phi+\lambda_2 B dt) + \frac{a^2x^2\mathbb{M}+ \mathbb{L}r^2\Omega^2(1-x^2)P }{r^4(1-x^2)P-a^2x^4Q}\left(\lambda_1 dt - \frac{2a (d\phi+\lambda_2 B dt)}{(1+\sqrt{I_2}) P_0}\right)\right\}^2\nonumber\\
&+\frac{(1-x^2)P}{\Omega^4\mathbb{L}}\left\{\frac{r^2}{P_0}(d\phi+\lambda_2 B dt)+ \frac{a\,\left(r^2\mathbb{M}+ \mathbb{L}Q\Omega^2x^2\right)}{r^4(1-x^2)P-a^2x^4Q}\left(\lambda_1 dt  - \frac{2a (d\phi+\lambda_2 B dt)}{(1+\sqrt{I_2}) P_0} \right)\right\}^2\nonumber\\
&+\mathbb{L}\left(\frac{dr^2}{Q}+\frac{dx^2}{(1-x^2)P}\right)\,.
\end{align}
\end{widetext}
The metric functions appearing in Eq.~\eqref{eq:metric3} are the same as those given in Eqs.~\eqref{eq:MLfunctions}--\eqref{eq:P0I1I2}. With the thermodynamic normalization, the conserved mass and angular momentum are
\bea
M
&=&
\frac{
\gamma(\varepsilon_2-\varepsilon_1)
\sqrt{(\varepsilon_1+1)(\varepsilon_2^2-1)}
}{
2\varepsilon_2^{3/2}
(\varepsilon_1^2\varepsilon_2-2\varepsilon_1+\varepsilon_2)^{3/2}
B
} \,, \label{eq:Mdef}\\
\quad
J
&=&
\frac{
(\varepsilon_2-\varepsilon_1)(\varepsilon_1+1)^2
\sqrt{1-\varepsilon_1^2}
(\varepsilon_2^2-1)^{3/2}
}{
4\varepsilon_2^2
(\varepsilon_1^2\varepsilon_2-2\varepsilon_1+\varepsilon_2)^2
B^2
} \,,\label{eq:Jdef}
\eea
where
\bea
\gamma &=&
\sqrt{
(1+\varepsilon_1+\varepsilon_1^2-\varepsilon_1^3)\varepsilon_2^2
+2(\varepsilon_1-1)\varepsilon_1\varepsilon_2
-2\varepsilon_1
}\,, \nn\\
\varepsilon_1 &=&\sqrt{I_2} \,,\quad
\varepsilon_2= \sqrt{ 1+B^2r_+^2}   \,.  
\eea
The constants $\lambda_1$ and $\lambda_2$ are fixed as
\begin{align}\label{eq:lambdadef}
\lambda_1
&=
\frac{
2\varepsilon_1(\varepsilon_1+1)(\varepsilon_2^2-1)-\gamma^2
}{
\gamma
\sqrt{
\varepsilon_2(\varepsilon_1+1)(\varepsilon_1^2\varepsilon_2-2\varepsilon_1+\varepsilon_2)
}
} \,, \nn \\
\lambda_2
&=
\frac{
\left[
2\varepsilon_1\varepsilon_2+2\varepsilon_1-(\varepsilon_1+1)^2\varepsilon_2^2
\right]\sqrt{
\varepsilon_2(1-\varepsilon_1)(\varepsilon_1^2\varepsilon_2-2\varepsilon_1+\varepsilon_2)
}
}{
(\varepsilon_1+1)(\varepsilon_2^2-1)\gamma
} \,.
\end{align}
With this choice, the conserved charges $M$ and $J$, together with the temperature $T$, entropy $S$, and angular velocity $\omega$ of the horizon, satisfy both the differential and integral forms of the first law of black-hole thermodynamics,
\be
d M = T d S + \omega d J \,, \quad M = 2 T S + 2 \omega J \,. \nn
\ee
The thermodynamic analysis therefore identifies the physical mass and angular momentum of the ML black hole. This is particularly important for phenomenological studies, since the integration constants appearing in a metric need not coincide directly with the physical conserved charges, and using them as phenomenological parameters may obscure or even misidentify the physical interpretation of observational constraints~\cite{Yu:2025odj,Luo:2026duf}. We therefore characterize the geodesic and shadow observables in terms of the physical quantities $M$ and $J$, rather than solely in terms of the integration constants $m$ and $a$.
It is then more appropriate to characterize the rotation in terms of the dimensionless spin parameter
 $\chi$
 \be
 \chi \equiv \frac{J}{M^2} = \frac{(\varepsilon_1 +1) \varepsilon_2  \left(\varepsilon_1 ^2 \varepsilon_2 -2 \varepsilon_1 +\varepsilon_2 \right)  }{\gamma^2 (\varepsilon_2 -\varepsilon_1 )} \sqrt{(1-\varepsilon_1^2)(\varepsilon_2^2-1)} \,. \label{eq:chidef}
 \ee
The metric~\eqref{eq:metric3}, together with the physical mass $M$~\eqref{eq:Mdef} and dimensionless spin $\chi$~\eqref{eq:chidef}, provides the background for the geodesic and shadow analysis below. Throughout both the analytic and numerical calculations, we characterize the black hole by the physical parameters $M$, $\chi$, and $B$, which enables physically meaningful comparisons with the Kerr spacetime at fixed mass and spin. Because the exact metric is rather involved, a closed analytic treatment is difficult in general. We therefore use a small-deformation expansion, $B\ll1$, for the analytic part of the discussion, while the full expressions are retained in the numerical analysis. In this regime, the integration constants $m$ and $a$ can be solved perturbatively in terms of the physical parameters $M$ and $\chi$. Solving Eqs.~\eqref{eq:Mdef} and \eqref{eq:chidef} order by order in $B$, we obtain, up to order ${\cal O}(B^2)$,
\bea
m &=& M +\frac{B^2 M^3}{4}  \left(6 -5 \chi^2\right)  \,,\\
a &=&  M \chi +\frac{B^2 M^3 \chi }{4}   \left(2 -\chi^2\right)  \,.
\eea
It is useful to note the dimension of the spindle parameter $B$. Since $B$ always appears through dimensionless combinations such as $Br$, and $B M$, it has the dimension of inverse length in geometrized units, $[B]=M^{-1} $.
After the solution is expressed in terms of the physical mass $M$, a perturbative expansion in the spindle parameter $B$ should be understood as an expansion in the dimensionless combination $BM\ll1$. Substituting the perturbative relations into the extremality condition~\eqref{extreme}, we find that, up to order $\mathcal{O}(B^2)$, the extremal limit is still given by $\chi=1$, as in Kerr. Therefore, at this perturbative order, the black-hole condition corresponds to $|\chi|\leq 1$. The corresponding horizon radius is shifted to
\be
r_+^{\rm extreme}
=
M+\frac{3}{4}B^2M^3 \,. \label{extremehorizon}
\ee

\section{Geodesics}\label{geodesicsection}

We now study timelike and null geodesics in the ML black hole described by the metric~\eqref{eq:metric3}, with the physical mass and angular momentum given by Eqs.~\eqref{eq:Mdef} and \eqref{eq:Jdef}, respectively. Since the analytic treatment of geodesic motion is closely tied to the separability of the Hamilton-Jacobi equation, we first examine the separability structure of the spacetime. The geodesic structure of the Kerr spacetime has been extensively studied since the classic works on separability and black-hole particle dynamics~\cite{Carter:1968rr,Chandrasekhar:1985kt} (see, e.g., Refs.~\cite{Walker:1970un,Wilkins:1972rs,Bardeen:1972fi,Teo:2003ltt,Gralla:2019ceu,Bardeen:1973tla,Johannsen:2013vgc,Compere:2021bkk,Liu:2023tcy,Cieslik:2023qdc} for further studies of geodesics in Kerr spacetime). For the ML black hole, however, such a hidden symmetry cannot be assumed a priori, because the spindle deformation changes both the Kerr geometry and its asymptotic structure. We therefore analyze the Hamilton-Jacobi equation explicitly. This analysis also determines how far analytic methods can be used for circular orbits, photon regions, and black-hole shadows.

\subsection{Hamilton-Jacobi equation and separability}

The Hamilton-Jacobi equation for a freely moving particle is given by
\be
\frac{\partial S}{\partial \tau}
+\frac12 g^{\mu\nu}
\frac{\partial S}{\partial x^\mu}
\frac{\partial S}{\partial x^\nu}
=0 \,,
\label{HJeq}
\ee
where $S$ is Hamilton's principal function and $\tau$ is an affine parameter along the geodesic. For the metric~\eqref{eq:metric3}, this equation takes the form
\be
0
=
2\frac{\partial S}{\partial \tau}
+g^{tt}\left(\frac{\partial S}{\partial t}\right)^2
+2g^{t\phi}
\frac{\partial S}{\partial t}
\frac{\partial S}{\partial \phi}
+g^{\phi\phi}
\left(\frac{\partial S}{\partial \phi}\right)^2
+g^{rr}
\left(\frac{\partial S}{\partial r}\right)^2
+g^{xx}
\left(\frac{\partial S}{\partial x}\right)^2 \,.
\label{HJeq2}
\ee
To examine separability, we adopt the standard ansatz
\be
S =
\frac12 \mu^2 \tau
-Et
+L\phi
+S_r(r)
+S_x(x) \,,
\label{principalf}
\ee
where $E$ and $L$ are the conserved energy and angular momentum of the particle associated with the Killing vectors $\partial_t$ and $\partial_\phi$, respectively. The parameter $\mu$ denotes the particle mass, with $\mu=1$ for massive particles and $\mu=0$ for massless particles. Substituting Eq.~\eqref{principalf} into Eq.~\eqref{HJeq2}, and using $g^{rr}=Q/\mathbb{L}$ and $g^{xx}=P(1-x^2)/\mathbb{L}$, we obtain
\be
\frac{1}{\mathbb{L}}
\left[
\mathbb{L} \mu^2
+ X
+ Q \left(\frac{d S_r}{d r}\right)^2
+ P(1-x^2) \left(\frac{d S_x}{d x}\right)^2
\right]
=0\,, \label{separate}
\ee
where
\be
X =
\mathbb{L}
\left(
g^{tt}E^2
-2g^{t\phi}EL
+g^{\phi\phi}L^2
\right) \,.
\ee
In the full geometry, the metric functions entering Eq.~\eqref{separate} contain nontrivial mixed dependence on $r$ and $x$. Consequently, the Hamilton-Jacobi equation is not additively separable in the Boyer-Lindquist-type coordinates used here. To obtain analytic control, we therefore expand the Hamilton-Jacobi equation in powers of the spindle parameter $B$, after expressing the metric in terms of the physical mass $M$ and the dimensionless spin parameter $\chi$. Up to order ${\cal O}(B^2)$, the function $\mathbb{L}$ is
\bea
\mathbb{L} &=& M^2 \chi ^2 x^2+r^2 + B^2 \Bigg[\frac{3}{2} M^4 \chi ^4 x^4+\frac{3}{2} M^2 r^2 \chi ^2 x^4 
-\frac{1}{2} M^4 \chi ^2 \left(\chi ^2-2\right) x^2-3 M^3 r \chi ^2 x^4 \nn\\
&\quad&-2 M r^3 x^2+r^4 \left(\frac{3 x^2}{2}-\frac{3}{2}\right)\Bigg] \,.
\eea
In the Kerr limit, $B=0$, the term $\mathbb{L}\mu^2$ separates additively into a radial part and an angular part. Once the spindle deformation is included, however, $\mathbb{L}\mu^2$ contains mixed dependence on $r$ and $x$. This obstruction persists even in the static limit $\chi=0$, for which the $B^2$ correction contains mixed terms such as $-2Mr^3x^2$. Therefore, the timelike Hamilton-Jacobi equation is not additively separable at order ${\cal O}(B^2)$.

The null sector is different. For $\mu=0$, the nonseparable term $\mathbb{L}\mu^2$ drops out. Although the remaining quantity $X$ has a lengthy expression, its expansion up to order ${\cal O}(B^2)$ satisfies
\be
\frac{\partial^2 X}{\partial r \partial x}=0 \,.
\ee
Therefore, at this order, $X$ can be written as
\be
X = X_1(r) +X_2(x) \,.
\ee
The explicit expressions of $X_1(r)$ and $X_2(x)$ are
\begin{widetext}
\bea
X_1(r) &=& 
\frac{r}{2\left(-2Mr+r^2+M^2\chi^2\right)^2}
\Big\{
-2\left(-2Mr+r^2+M^2\chi^2\right)
\Big[
L^2\left(2M-r\right)
\nonumber\\
&\quad&
-4ELM^2\chi+E^2\left(r^3+2M^3\chi^2+M^2r\chi^2\right)
\Big]+B^2
\Big[
L^2
\Big\{
8Mr^4
-2r^5
\nonumber\\
&\quad&
+5M^5\chi^2\left(-2+\chi^2\right)+4M^3r^2\left(-4+3\chi^2\right)-2M^2r^3\left(2+3\chi^2\right)
\nonumber\\
&\quad&
+M^4r\left(16-3\chi^4\right)
\Big\}
+E^2
\Big\{
-4Mr^6
+2r^7
+M^7\chi^4\left(-10+\chi^2\right)
\nonumber\\
&\quad&
+M^2r^5\left(-4+\chi^2\right)-4M^4r^3\chi^2\left(4+\chi^2\right)
+6M^5r^2\chi^2\left(-2+3\chi^2\right)
\nonumber\\
&\quad&
+M^3r^4\left(-2+13\chi^2\right)
+M^6r\chi^2\left(16+4\chi^2-3\chi^4\right)
\Big\}
\nonumber\\
&\quad&
+2ELM\chi
\Big\{
2r^5-8Mr^4
+M^3r^2\left(14-13\chi^2\right)
+2M^2r^3\left(5+2\chi^2\right)
\nonumber\\
&\quad&
+M^5\left(10\chi^2-3\chi^4\right)
+2M^4r\left(\chi^4-8-\chi^2\right)
\Big\}
\Big]
\Big\} \,, \label{X1equation} \\
 X_2(x) &=& x^2
\Bigg\{
2B^2 E L M^3 \chi \left(1+\chi^2\right)
+\frac{
L^2\left[1+B^2M^2\left(1+\left(-2+x^2\right)\chi^2\right)\right]
}
{1-x^2} \nn \\
&\quad&
-E^2M^2\chi^2
\left[
1+B^2M^2\left(3+\chi^2+x^2\left(-1+\chi^2\right)\right)
\right]
\Bigg\} \,. \label{X2equation} 
\eea
\end{widetext}
However, this separability does not persist at the next nontrivial order. A direct expansion to ${\cal O}(B^4)$ reveals mixed $r$- and $x$-dependent contributions, with the ${\cal O}(B^4)$ term satisfying $\frac{\partial^2 X(r,x)}{\partial r \partial x}\neq 0$. Therefore, the appearance of nonseparable contributions at order ${\cal O}(B^4)$ demonstrates that no exact separation constant exists for the full null Hamilton-Jacobi equation. This result shows that the timelike Hamilton-Jacobi equation is already nonseparable at order ${\cal O}(B^2)$, whereas the null Hamilton-Jacobi equation becomes separable at the same perturbative order. This should be contrasted with the KBR seed geometry, for which the null Hamilton-Jacobi equation is exactly separable~\cite{Wang:2025vsx}. In the analytic treatment below, we therefore restrict timelike circular motion to the equatorial plane, while the perturbatively separable null sector allows us to characterize the photon region beyond the equatorial plane.

\subsection{Timelike geodesics and stable circular orbits} \label{timeg}

\begin{figure*}[t]
    \centering
    \includegraphics[width=0.9\textwidth]{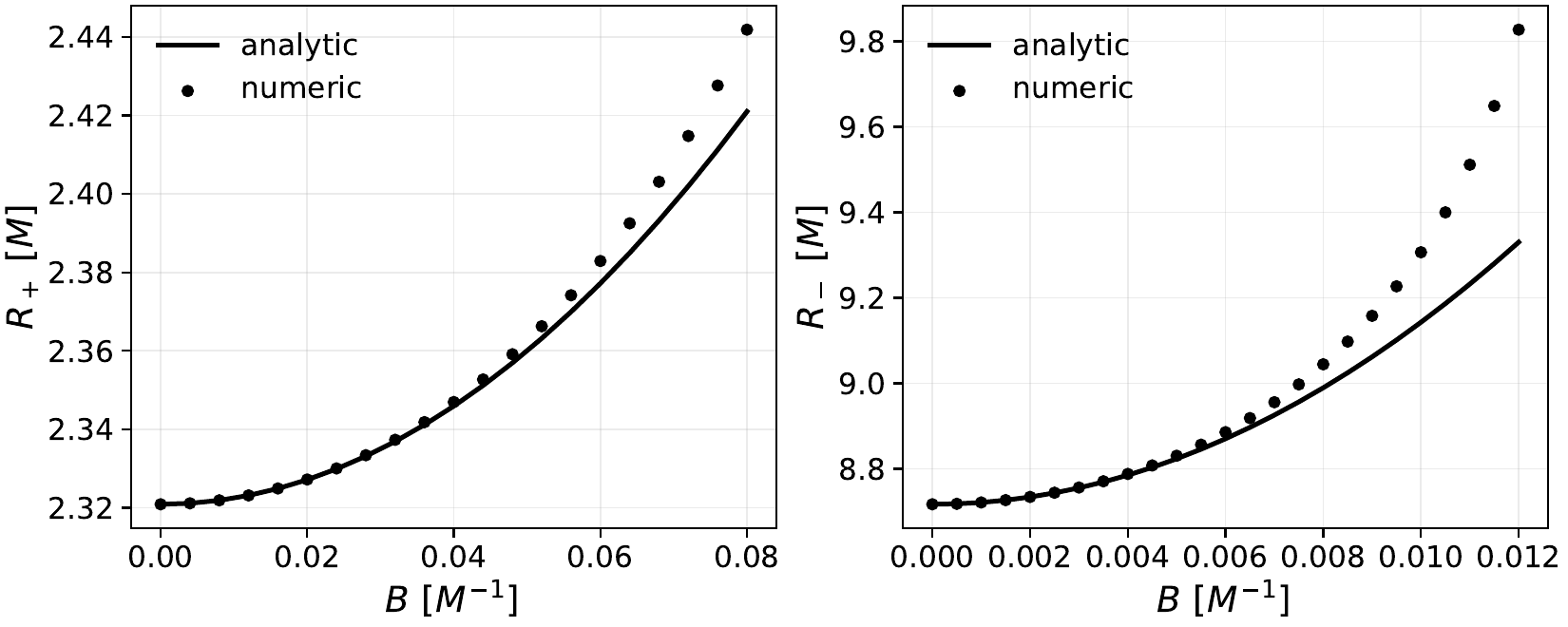}
\caption{Comparison between the numerical and perturbative ISCO radii as functions of the dimensionless spindle parameter $BM$ at fixed spin parameter $\chi=0.9$. The left and right panels show the corotating and counterrotating branches, respectively. In each panel, the points are obtained from the full effective potential~\eqref{potential1}, while the solid curves denote the perturbative result based on Eq.~\eqref{pertisco}.}
    \label{fig:isco_physical_numeric}
\end{figure*}

Since the Hamilton-Jacobi equation for generic timelike geodesics is not separable already at order ${\cal O}(B^2)$, we restrict the analytic discussion of timelike motion to equatorial circular geodesics. The metric functions are even under $x\to -x$, so the equatorial plane $x=0$ is invariant under the geodesic flow: initial data with $x=0$ and $d{x}/d\tau=0$ remain confined to this plane. For timelike geodesics, we set $\mu=1$. The canonical momentum is related to the four-velocity by
\be
p_\mu=\frac{\partial S}{\partial x^\mu}
=g_{\mu\nu}\frac{dx^\nu}{d\tau} \,. \label{pmu}
\ee
Using Eq.~\eqref{separate}, we obtain the radial equation on the equatorial plane,
\be
V(r) \equiv \left(\frac{ dr}{d\tau}\right)^2
= - \frac{Q}{\mathbb{L}^2}
\left(\mathbb{L}+X\right)\Big|_{x=0} \,. \label{potential1}
\ee
Marginally stable circular orbits are determined by
\be
V(R) = 0 \,, \quad \frac{dV(R)}{dr} = 0  \,, \quad \frac{d^2V(R)}{dr^2} = 0 \,. \label{ISCOeq}
\ee
Here, $R$ denotes the radius of a marginally stable circular orbit. Among the physically admissible solutions outside the outer horizon, the innermost one defines the ISCO. As will be seen below, for finite spindle deformation an additional outer marginally stable circular orbit may also appear; we refer to it as the outermost stable circular orbit (OSCO). Before performing the full numerical analysis, we first derive an analytic expression for the ISCO radius in the small-deformation regime. Expanding the radial potential $V$ in powers of the spindle parameter $B$, we obtain, up to order ${\cal O}(B^2)$,
\begin{widetext}
\bea\label{potential2}
V &=&
\frac{1}{r^3}
\Big\{2E^2M^3\chi^2
+\left(E^2-1\right)M^2r\chi^2
+\left(E^2-1\right)r^3
-4ELM^2\chi
+L^2\left(2M-r\right)
 \nn\\
&\quad& +2Mr^2\Big\} +\frac{B^2}{2r^3}
\Big\{
E^2 M^5 \chi^2 \left(14-3\chi^2\right)
+M^4 r \chi^2
\left(
2E^2\left(\chi^2+3\right)+\chi^2-2
\right)
\nonumber\\
&\quad&
+M^3 r^2
\left[
\left(12E^2-7\right)\chi^2+6
\right]
+M^2 r^3
\left[
E^2\left(9\chi^2+2\right)-5\chi^2+2
\right]
+\left(6E^2-5\right)r^5
\nonumber\\
&\quad&
+2ELM\chi
\left[
M^3\left(5\chi^2-14\right)
-2M^2r\left(\chi^2+1\right)
-12Mr^2
-2r^3
\right]
\nonumber\\
&\quad&
+L^2
\left[
2M^2r\left(2\chi^2-1\right)
-7M^3\left(\chi^2-2\right)
+12Mr^2
-6r^3
\right]
+10Mr^4
\Big\} \,.
\eea
\end{widetext}
We assume that the ISCO radius admits the perturbative expansion
\be
R = R_{K} + B^2 M^2 R_{B} +\mathcal{O}(B^4) \,, \label{pertisco}
\ee
where the zeroth-order term is the Kerr ISCO radius,
\be
R_{K\pm}
=
M
\left[
3+Z_2
\mp
\sqrt{(3-Z_1)(3+Z_1+2Z_2)}
\right] \,,
\ee
with
\bea
Z_1
&=&
1+
\left(
1-\chi^2
\right)^{1/3}
\left[
\left(
1+\chi
\right)^{1/3}
+
\left(
1-\chi
\right)^{1/3}
\right] \,,
\nn\\
Z_2
&=&
\sqrt{
3\chi^2
+
Z_1^2
} \,.
\eea
Here and below, the upper sign denotes the corotating branch, while the lower sign denotes the counterrotating branch. The leading correction induced by the spindle deformation is
\begin{widetext}
\bea
R_{B\pm}
&=&
M \Big[
3 M^4 R_{K\pm}^3 \left(14 - 13 \chi^2\right)
+ 12 M^2 R_{K\pm}^5 \left( \chi^2-10 \right)
+ 7 M^6 R_{K\pm} \chi^2 \left(6 + 5 \chi^2\right)
\nn \\
&\quad&
+ 42 M R_{K\pm}^6 
+ M^3 R_{K\pm}^4 \left(66 + 41 \chi^2\right)
+ 3 M^5 R_{K\pm}^2
\left( \chi^4 -12 - 22 \chi^2 \right)
- 4 R_{K\pm}^7
\nn \\
&\quad&
- 18 M^7 \chi^4
\Big]\Big/\Big[
4 M^4
\left(
9 M R_{K\pm}^2
- R_{K\pm}^3
+ 7 M^3 \chi^2
+ 3 M^2 R_{K\pm} \left( \chi^2-6 \right)
\right)
\Big] \,.
\eea
\end{widetext}
It is useful to examine several limiting cases. First, when $B=0$, the standard Kerr result is recovered. Second, in the static limit $\chi=0$, the two branches coincide, and the perturbative result becomes
\be
R^{\rm static} = 6 M + 879 B^2 M^3 \,.
\ee
This limit isolates the effect of the spindle deformation in the absence of rotation. Third, we consider the extremal rotating limit. At order $B^2$, the extremality condition reduces to $\chi=1$, and the perturbative ISCO radius becomes
\bea
R_{+}^{\rm extreme} &=& M + \frac{3 B^2 M^3}{4} \,,\nn\\
R_{-}^{\rm extreme} &=& 9 M + \frac{19467 B^2 M^3}{4} \,. \label{extremeisco}
\eea
The perturbative result shows that the spindle deformation shifts both corotating and counterrotating ISCO radii outward. The effect is especially pronounced for the counterrotating branch in the rapidly rotating regime. We next perform a full numerical analysis of marginally stable circular orbits by solving Eq.~\eqref{ISCOeq} with the full effective potential~\eqref{potential1}. As an illustrative example, we fix the spin parameter to $\chi=0.9$ and compare the ISCO radii obtained from the full effective potential with those obtained from the perturbative potential~\eqref{potential2}. The comparison for the corotating and counterrotating branches is shown in Fig.~\ref{fig:isco_physical_numeric}.
The comparison shows that the perturbative result tracks the full numerical result for the corotating branch up to $BM\sim 10^{-2}$, whereas for the counterrotating branch visible deviations appear already around $BM\sim 10^{-3}$. This branch dependence is consistent with the extremal-limit expression~\eqref{extremeisco}: in the rapidly rotating regime, the counterrotating ISCO carries a much larger $B^2$ coefficient than the corotating one. Therefore, the counterrotating branch is more sensitive to the spindle deformation, which is consistent with the roughly one-order-of-magnitude smaller perturbative range observed in Fig.~\ref{fig:isco_physical_numeric}.

\begin{figure*}[t]
    \centering
    \includegraphics[width=0.9\textwidth]{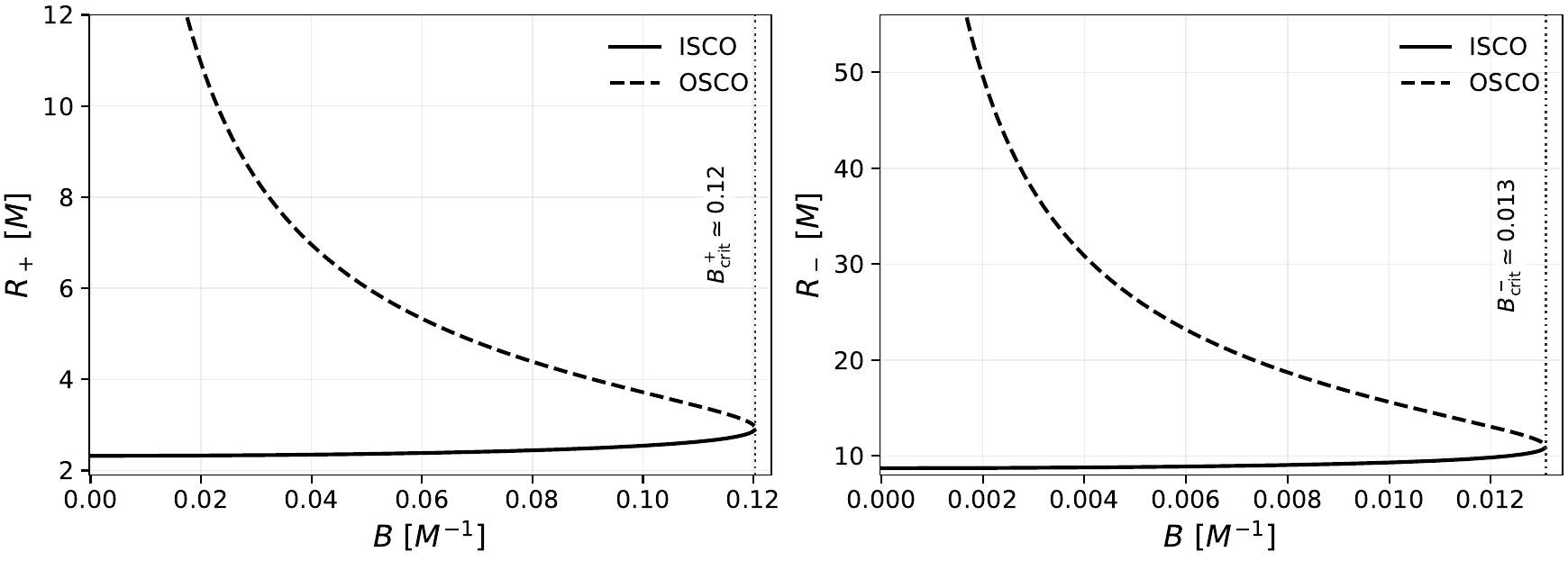}
    \caption{Numerical ISCO and OSCO radii as functions of the dimensionless spindle parameter $BM$ at fixed spin parameter $\chi=0.9$. The left and right panels show the corotating and counterrotating branches, respectively. In each panel, the stable circular-orbit region lies between the ISCO and OSCO curves, which merge at a critical value of $BM$.}
    \label{fig:isco_osco}
\end{figure*}

Besides testing the validity of the perturbative solution, the full numerical analysis reveals an additional feature that is absent in the local analytic expansion around the Kerr ISCO. For nonzero spindle deformation, the stable circular-orbit region need not extend to arbitrarily large radius. Instead, it can become a finite radial interval bounded by an inner marginally stable orbit, identified as the ISCO, and an outer marginally stable orbit, identified as the OSCO, a feature already observed in the static $B$-deformed black hole~\cite{Astorino:2026okd}. This behavior is shown in Fig.~\ref{fig:isco_osco}.
As $B$ increases, the ISCO radius increases, whereas the OSCO radius decreases. The two boundaries eventually merge at a critical value of $B$, above which no stable circular-orbit region is found for the corresponding branch. This behavior is reminiscent of other non-asymptotically flat black-hole spacetimes, such as Kerr-Melvin~\cite{Hou:2022eev} and Kerr-de Sitter~\cite{Boonserm:2019nqq} black holes, where stable circular orbits can also be bounded by an outer marginally stable orbit. In the present case, the OSCO is a genuinely finite-$B$ feature associated with the spindle deformation. It is not captured by the perturbative ISCO solution, because that solution is a local expansion around the Kerr ISCO, whereas the OSCO is controlled by the large-radius/global structure of the radial potential and moves to infinity as $B\to0$.

\subsection{Null geodesics and photon region}

Since the null Hamilton-Jacobi equation becomes separable through order ${\cal O}(B^2)$, the null geodesics of the perturbatively expanded system can be analyzed in close analogy with those of Kerr. Setting $\mu=0$, Eq.~\eqref{separate} separates into
\bea
X_1(r) + Q \left(\frac{d S_r}{d r}\right)^2 &=& - K \,,  \label{nullhj1} \\
X_2(x) + P(1-x^2) \left(\frac{d S_x}{d x}\right)^2 &=& K \,, \label{nullhj2}
\eea
where $K$ is a Carter-like separation constant. The functions $P$ and $Q$ are given in Eq.~\eqref{eq:MLfunctions}, while $X_1$ and $X_2$ are given in Eqs.~\eqref{X1equation} and \eqref{X2equation}, respectively. The radial and angular equations of motion can be written as
\bea
\mathbb{L} \frac{dr}{d\tau} &=& \frac{dr}{d\tilde{\tau}} = \pm \sqrt{ \Xi (r)} \,, \\
\mathbb{L} \frac{dx}{d\tau} &=& \frac{dx}{d\tilde{\tau}} =  \pm \sqrt{ \Theta (x)} \,, 
\eea
where the Mino-like parameter $\tilde{\tau}$~\cite{Mino:2003yg} is defined by $d \tilde{\tau} = d \tau/\mathbb{L}$. The radial and angular potentials are
\bea
\Xi(r) &=& {-Q(r) (K + X_1(r) )} \,,\\
 \Theta(x) &=& {P(x)(1-x^2) (K - X_2(x))} \,.
\eea
The projection of the photon trajectory onto the $(r,x)$ plane is therefore determined by
\be
\int_{r_i}^r \frac{dr}{\sqrt{\Xi}} = \int_{x_i}^x \frac{dx}{\sqrt{\Theta}} \,,
\ee
where $r_i$ and $x_i$ denote the initial values of $r$ and $x$, respectively. The signs of $dr/d\tilde{\tau}$ and $dx/d\tilde{\tau}$ can be chosen independently, but once chosen they must be kept fixed along a given monotonic segment of the trajectory. Since an overall rescaling of the photon energy does not change the trajectory, we introduce the dimensionless impact parameters
\be
\xi = \frac{L}{E} \,,\quad  \eta= \frac{ K}{E^2} \,.
\ee
In what follows, we rescale the potentials as $\Xi\to\Xi/E^2$ and $\Theta\to\Theta/E^2$.

We first analyze the angular motion. The angular potential takes the quartic form
\be
\Theta =  -A_1 x^4  - A_2 x^2 +\eta \,,
\ee
where
\bea
A_1 &=& M^2 \chi ^2 + B^2 M^2 \left[ \left(1-\chi ^2\right) (\eta  +M^2  \left(\chi ^2+2\right) \chi ^2)+\left(\xi -M \chi ^3-M \chi \right)^2\right] \,, \nn \\
A_2 &=& \eta +\xi ^2  -M^2 \chi ^2 - B^2 M^2 \big[\eta +M^2 \chi ^2 \left(\chi ^2+3\right) -\eta  \chi ^2 -2 M \xi  \chi  \left(\chi ^2+1\right) \nn \\
&\quad& +\xi ^2 \left(2 \chi ^2-1\right)\big] \,, \label{A1A2}
\eea
The allowed angular motion is determined by $\Theta(x)\ge0$. In particular,
\bea
\Theta (x =0) &=& \eta \,, \nn\\
\Theta (x =1) &=& -\xi ^2 \left(1 - 2 B^2 M^2 \left(1 -\chi ^2\right)\right) \,.
\eea
In the small-deformation regime $BM\ll1$, one has $\Theta(1)\le0$. As in Kerr, the sign of $\eta=\Theta(0)$ determines whether the photon trajectory can cross the equatorial plane. For $\eta>0$, the photon can cross the equatorial plane; for $\eta=0$, the equatorial plane is a limiting turning point; for $\eta<0$, the photon cannot cross the equatorial plane, although allowed off-equatorial intervals may still exist. 
Since the equatorial-crossing sector provides the critical null orbits used in the shadow construction considered below, we restrict the following analysis to $\eta\ge0$. In this case the angular potential can be factorized as
\be
\Theta = A_1 (x_-^2 + x^2) (x_+^2 - x^2)     \,,
\ee
where $A_1>0$ and 
\be
x_\pm^2 = \frac{\mp A_2 + \sqrt{A_2^2 + 4A_1\eta} }{2A_1} \,. \label{root1}
\ee
The allowed angular region is
\be
0\le x^2\le x_+^2 \,,
\quad \text{or} \quad
-x_+\le x\le x_+ \,.
\ee
Thus, for $\eta\ge0$, the photon trajectory can oscillate symmetrically across the equatorial plane.
The corresponding angular integral is
\be
\int_x^{x_+} \frac{dx}{\sqrt{ \Theta}} = \frac{F\left(\cos^{-1}(\frac{x}{x_+} )\,, \frac{x_+}{\sqrt{x_+^2 + x_-^2}}\right) }{\sqrt{A_1 (x_+^2 + x_-^2)} }  \,,
\ee
where $F$ represents the incomplete elliptic integral of the first kind. For small spindle deformation, the angular turning point $x_+$ receives a correction. Substituting Eq.~\eqref{A1A2} into Eq.~\eqref{root1} and expanding $x_+^2$ in powers of $B$, we obtain
\be
x_+^2 = x_{K+}^2 + x_{B+}^2 B^2 \,,
\ee
where the Kerr result is 
\be
x_{K+}^2 = \frac{M^2 \chi ^2-\xi ^2-\eta +\sqrt{\eta ^2+2 \eta  \left(M^2 \chi ^2+\xi ^2\right)+\left(\xi ^2-M^2 \chi ^2\right)^2}}{2 M^2 \chi ^2}\,.
\ee
The explicit expression of $x_{B+}^2$ is given by
\begin{widetext}
\bea
x_{B+}^2 &= &
\frac{1}{2 M^2\chi^2}\Bigg\{-\frac{1}{\chi^{2}}
\left[
-\eta\chi^2+\eta
+M^2\chi^2\left(\chi^2+3\right)
-2M\xi\chi\left(\chi^2+1\right)
+\xi^2
\right] \nn\\
&\quad& \left[
-\eta
+\sqrt{
\eta^2
+2\eta\left(M^2\chi^2+\xi^2\right)
+\left(\xi^2-M^2\chi^2\right)^2
}
+M^2\chi^2-\xi^2
\right] \nn \\
&\quad&
+M^2
\left[
-\eta\chi^2+\eta
+M^2\chi^2\left(\chi^2+3\right)
-2M\xi\chi\left(\chi^2+1\right)
+\xi^2\left(2\chi^2-1\right)
\right] \nn \\
&\quad& +\frac{M^2}{\big(
\eta^2
+2\eta\left(M^2\chi^2+\xi^2\right)
+\left(\xi^2-M^2\chi^2\right)^2
\big)^{\frac12} }
\Big[
M^4\chi^4\left(\chi^2+3\right) \nn \\
&\quad&
-\eta^2\left(\chi^2-1\right)
-2M^3\xi\chi^3\left(\chi^2+1\right)
-\eta
\big(
+2M\xi\chi\left(\chi^2+1\right)
-4M^2\chi^2\nn \\
&\quad&
+\xi^2\left(\chi^2-2\right)
\big)
+M^2\xi^2\chi^2\left(\chi^2-4\right)
+2M\xi^3\chi\left(\chi^2+1\right)
+\xi^4\left(1-2\chi^2\right)
\Big]
\Bigg\} \,.
\eea
\end{widetext}

We now turn to the radial motion. Within the perturbatively separated null dynamics, the critical trajectories generating the shadow boundary are associated with unstable null orbits of constant radial coordinate. Denoting the radius of such an orbit by ${\cal R}$, the corresponding conditions are
\be
\Xi ({\cal R}) = 0 \,, \quad \frac{d\Xi ({\cal R}) }{dr} = 0 \,.
\ee
Solving above equation gives the critical impact parameters $\xi$ and $\eta$ as functions of the photon-orbit radius ${\cal R}$:
\begin{widetext}
\bea
\eta &=& \frac{
{\cal R}^3\left(
-9M^2{\cal R}
+6M{\cal R}^2
-{\cal R}^3
+4M^3\chi^2
\right)
}
{
M^2\left(M-{\cal R}\right)^2\chi^2
}+\frac{
B^2 {\cal R}^3
}{
M^4\left(M-{\cal R}\right)^3\chi^4
}
\Big[
M^2{\cal R}^6\left(45+2\chi^2\right)
\nonumber\\
&\quad&
+{\cal R}^8-11M{\cal R}^7
+M^8\chi^4\left(-2+5\chi^2\right)-M^7{\cal R}\chi^4\left(18+7\chi^2\right)
-3M^3{\cal R}^5\left(27+8\chi^2\right)
\nonumber\\
&\quad&
-M^5{\cal R}^3\chi^2\left(106+25\chi^2\right)
+M^6{\cal R}^2\chi^2\left(30+47\chi^2\right)
+2M^4{\cal R}^4
\left(
27+43\chi^2+2\chi^4
\right)
\Big] \,, \label{etaequation}\\
\xi &=& \frac{
-3M{\cal R}^2
+{\cal R}^3
+M^3\chi^2
+M^2{\cal R}\chi^2
}
{
M\left(M-{\cal R}\right)\chi
}+\frac{
B^2
}{
2M^3\left(M-{\cal R}\right)^2\chi^3
}
\Big[
8M{\cal R}^7
-{\cal R}^8
+M^8\chi^6
\nonumber\\
&\quad&
+3M^7{\cal R}\chi^4\left(-2+\chi^2\right)
-3M^2{\cal R}^6\left(7+\chi^2\right)
-M^4{\cal R}^4\chi^2\left(33+8\chi^2\right)
\nonumber\\
&\quad&
+2M^3{\cal R}^5\left(9+11\chi^2\right)
+M^5{\cal R}^3\chi^2\left(2+21\chi^2\right)
+M^6{\cal R}^2\chi^2
\left(6-4\chi^2-5\chi^4\right)
\Big] \,.\label{xiequation}
\eea
\end{widetext}
At the perturbative order considered here, the photon region is not characterized by a single radius. Rather, it is formed by the constant-$r$ null orbits of the separated system whose impact parameters $\xi({\cal R})$ and $\eta({\cal R})$ are compatible with the angular condition $\Theta(x)\ge0$. The equatorial photon orbits correspond to the limiting case $\eta=0$. For $\eta=0$, one has $\Theta(0)=0$. Since the angular potential is even in $x$, the initial data $x=0$ and $dx/d\tilde{\tau}=0$ define an equatorial null orbit. We therefore use $\eta=0$ to extract the equatorial photon-orbit radii. Assuming the equatorial photon-orbit radius admits the perturbative expansion
\be
{\cal R} = {\cal R}_{K} + B^2 M^2 {\cal R}_{B} +\mathcal{O}(B^4) \,, 
\ee
and substituting Eq.~\eqref{etaequation}, the zeroth-order term is the Kerr equatorial photon-orbit radius
\be
{\cal R}_{K\pm} = 2 M \left[ 1+\cos\left( \frac23 \cos^{-1}(\mp \chi) \right) \right] \,.
\ee
The upper sign denotes the corotating photon orbit, while the lower sign denotes the counterrotating photon orbit. The leading correction induced by the spindle deformation is
\begin{widetext}
\bea
{\cal R}_{B\pm}
&=&
-{\cal R}_{K\pm}
\Big[
4M^2\left(3M-{\cal R}_{K\pm}\right)\chi^2
\left(
-3M^2{\cal R}_{K\pm}
+3M{\cal R}_{K\pm}^{2}
-{\cal R}_{K\pm}^{3}
+M^3\chi^2
\right)
\Big]^{-1}
\nn\\
&&
\Big\{
-11 M {\cal R}_{K\pm}^{7}
+{\cal R}_{K\pm}^{8}
+M^2{\cal R}_{K\pm}^{6}\left(45+2\chi^2\right)
+M^8\chi^4\left(-2+5\chi^2\right)
\nn\\
&&
-M^7{\cal R}_{K\pm}\chi^4\left(18+7\chi^2\right)
-3M^3{\cal R}_{K\pm}^{5}\left(27+8\chi^2\right)
-M^5{\cal R}_{K\pm}^{3}\chi^2\left(106+25\chi^2\right)
\nn\\
&&
+M^6{\cal R}_{K\pm}^{2}\chi^2\left(30+47\chi^2\right)
+2M^4{\cal R}_{K\pm}^{4}\left(27+43\chi^2+2\chi^4\right)
\Big\} \,.
\eea
\end{widetext}
Although Eqs.~\eqref{etaequation} and \eqref{xiequation} are written for $\chi\neq0$, their apparent singular behavior at $\chi=0$ arises from the degeneracy of solving simultaneously for $\xi$ and $\eta$ in the static limit. The static photon-orbit radius can instead be obtained directly from the degenerate $\chi=0$ radial equations, or equivalently from the regular limit of the photon-orbit condition. In this limit, the equatorial photon-orbit radius is
\be
{\cal R}^{\rm static}
=
3M
+
\frac{15}{2}B^2M^3 \,.
\ee
At order $B^2$, the extremality condition reduces to $\chi=1$, and the perturbative equatorial photon-orbit radii become
\bea
{\cal R}_{+}^{\rm extreme} &=& M+\frac{3}{4}B^2M^3  \,,\nn\\
 {\cal R}_{-}^{\rm extreme} &=& 4M+3B^2M^3 \,. \label{extremephoton}
\eea
Thus, at order ${\cal O}(B^2)$, the spindle deformation increases the equatorial photon-orbit radius for both the corotating and counterrotating branches. In the extremal corotating limit, the radial-coordinate shift agrees with that of the horizon radius in Eq.~\eqref{extremehorizon}. Accordingly, the limiting corotating photon orbit remains coincident with the perturbed extremal horizon in the Boyer-Lindquist-type radial coordinate and at the perturbative order considered here. This behavior parallels the outward shift of the ISCO found in the timelike sector.

\section{Black-hole shadows}\label{shadowsection}

We now study the black-hole shadow of the ML black hole. We first construct the shadow boundary using the photon-region results obtained in the previous section. In the perturbative separable null sector, the critical photon orbits are characterized by the impact parameters $\xi({\cal R})$ and $\eta({\cal R})$ given in Eqs.~\eqref{etaequation} and \eqref{xiequation}. The shadow boundary is obtained by projecting this one-parameter family of critical photon momenta onto the local sky of an observer. We define the shadow at a finite observer radius using a local orthonormal tetrad, which allows a direct comparison with the Kerr shadow at the same physical mass $M$, spin $\chi$, observer radius $r_o$, and inclination angle $x=x_o=\cos\theta_o$. For a stationary and axisymmetric metric of the form~\eqref{eq:metric3}, a convenient local frame is the locally nonrotating frame~\cite{Bardeen:1973tla,Johannsen:2013vgc}. Its angular velocity and lapse function are
\be
\omega_o=-\frac{g_{t\phi}}{g_{\phi\phi}}\bigg|_o\,,
\quad
N_o=
\sqrt{
\frac{g_{t\phi}^2-g_{tt}g_{\phi\phi}}
{g_{\phi\phi}}
}\bigg|_o\,,
\label{zamoomega}
\ee
where the subscript $o$ indicates that the metric components are evaluated at the observer. The tetrad vectors are chosen as
\bea
e_{(t)}{}^\mu
&=&
\frac{1}{N_o}
\left(1,0,0,\omega_o\right)\,,\nn
 \\
e_{(r)}{}^\mu
&=&
\frac{1}{\sqrt{g_{rr}}}\delta_r{}^\mu\,,\nn
\\
e_{(x)}{}^\mu
&=&
\frac{1}{\sqrt{g_{xx}}}\delta_x{}^\mu\,,\nn
\\
e_{(\phi)}{}^\mu
&=&
\frac{1}{\sqrt{g_{\phi\phi}}}\delta_\phi{}^\mu\,.
\label{zamotetrad}
\eea
For a photon with covariant momentum
\be
p_\mu
=E\left(-1, \frac{p_r}{E}, \frac{p_x}{E}, \xi\right)\,,
\label{photonmomentum}
\ee
the locally measured spatial components are
\be
p_{(r)}=p_\mu e_{(r)}{}^\mu\,,
\quad
p_{(x)}=p_\mu e_{(x)}{}^\mu\,,
\quad
p_{(\phi)}=p_\mu e_{(\phi)}{}^\mu\,.
\label{localmomenta}
\ee
The screen coordinates $(\alpha, \beta)$ are defined by central projection with respect to the local radial direction,
\be
\alpha
=
r_o\frac{p^{(\phi)}}{p^{(r)}}\,,
\quad
\beta
=
-r_o\frac{p^{(x)}}{p^{(r)}}\,.
\label{screencoords}
\ee
The factor $r_o$ fixes the length scale used on the local observer screen. 
For a critical photon orbit with a radius ${\cal R}$, the ratios $p_r/E$ and $p_x/E$ are obtained from the separated radial and angular equations, with $\xi=\xi({\cal R})$ and $\eta=\eta({\cal R})$. The allowed part of the photon region is selected by the condition
\be
\Theta(x_o;\xi({\cal R}),\eta({\cal R}))\ge0\,.
\label{shadowangularcondition}
\ee
Thus the analytic photon-region result directly gives a parametric representation of the shadow boundary. For each allowed photon-orbit radius ${\cal R}$, substituting $\xi({\cal R})$ and $\eta({\cal R})$ into the separated equations gives the photon momentum at the observer,
\bea
\frac{p_r}{E}
&=&
\pm
\sqrt{
\frac{-\eta({\cal R})-X_1(r_o;\xi({\cal R}))/E^2}
{Q(r_o)}
}\,,
\nn\\
\frac{p_x}{E}
&=&
\pm
\sqrt{
\frac{\eta({\cal R})-X_2(x_o;\xi({\cal R}))/E^2}
{P(x_o)(1-x_o^2)}
}\,.
\label{observermomentum}
\eea
Here $X_1$ and $X_2$ are given by Eqs.~\eqref{X1equation}--\eqref{X2equation}. The radial sign is chosen so that the photon reaches the observer from the black hole, while the two signs of $p_x$ give the upper and lower halves of the image. After projection onto the local screen, they give the upper and lower parts of the image. Substituting Eq.~\eqref{observermomentum} into the local tetrad components~\eqref{localmomenta}, and then into Eq.~\eqref{screencoords}, gives the parametric shadow boundary
\be
\alpha=\alpha({\cal R})\,,
\quad
\beta=\beta_\pm({\cal R})\,.
\label{analyticshadowparam}
\ee
The sign in $\beta_\pm$ labels the two angular branches associated with $p_x$, and should not be confused with the corotating and counterrotating photon orbits, which are encoded in the behavior of the impact parameter $\xi({\cal R})$ along the allowed photon-region branch. By scanning the allowed range of ${\cal R}$, Eq.~\eqref{analyticshadowparam} gives a closed curve on the observer's screen, which we identify as the perturbative analytic shadow boundary.

\begin{figure}[t]
    \centering
    \includegraphics[width=0.4\textwidth]{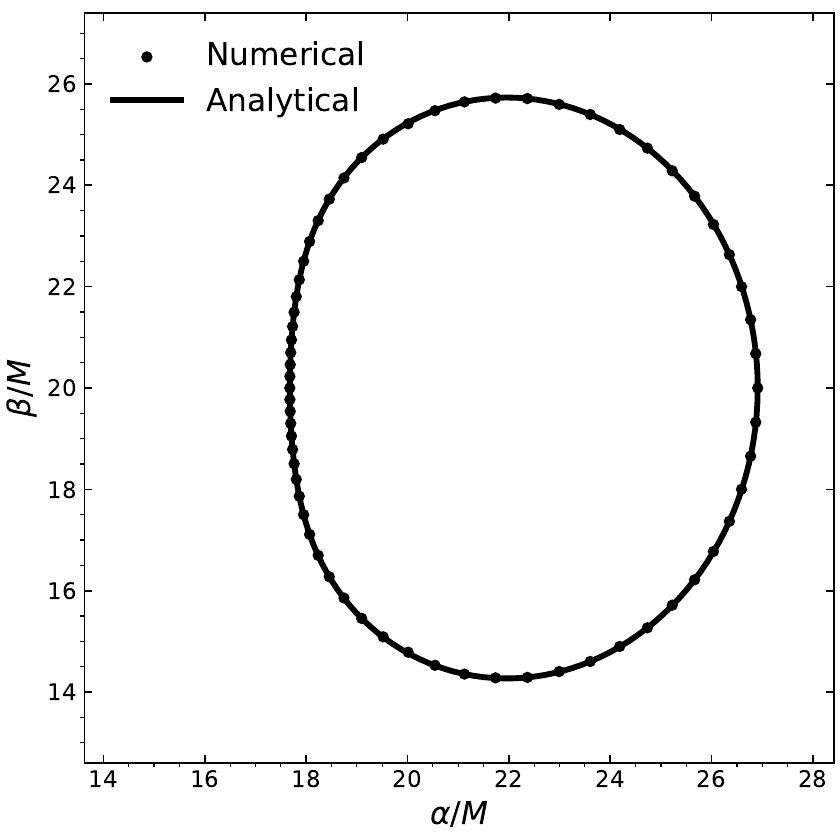}
    \caption{Comparison between the perturbative analytic shadow boundary and the full numerical ray-tracing result for $\chi=0.98$, $BM=0.01$, $r_o=50 M$, and $\theta_o=90^\circ$. The solid curve is obtained by projecting the perturbative photon region onto the observer screen, while the points are obtained from direct integration of the null geodesic equations.}
    \label{fig:shadow_validation}
\end{figure}

For comparison, we also compute the shadow by direct numerical ray tracing in the full metric~\eqref{eq:metric3}. For the numerical ray tracing, we use the original Hamilton-Jacobi equation~\eqref{separate}, with $\mu=0$ and~\eqref{pmu}. The photon trajectory is rewritten as
\be
\frac{dx^\mu}{d\tau}
=
g^{\mu\nu}p_\nu\,,
\quad
\frac{dp_\mu}{d\tau}
=
-\frac12
\partial_\mu g^{\alpha\beta}
p_\alpha p_\beta\,.
\label{hamiltoneq}
\ee
We parametrize the observer screen by polar coordinates $(\varrho, \psi)$,
\be
\alpha=\varrho\cos\psi\,,
\quad
\beta=\varrho\sin\psi\,.
\label{screenpolar}
\ee
For numerical ray tracing, the central-projection relation \eqref{screencoords} is used in the inverse direction. For each screen point $(\alpha,\beta)$, or equivalently for each pair $(\varrho,\psi)$, we determine the corresponding local photon direction and select the inward radial branch. The overall normalization of the null momentum is irrelevant and is fixed by $p^{(t)}=1$. The local momentum is then converted to the coordinate momentum by the tetrad \eqref{zamotetrad}, and the full null geodesic equations of the unexpanded metric are integrated from the observer. A ray is classified as captured if it reaches $r<r_+ +\epsilon$, where $\epsilon$ is a small numerical tolerance outside the outer horizon, and as escaping if it reaches a sufficiently large outer radius. For each angle $\psi$, we locate the transition between captured and escaping rays by bisection in the screen radius $\varrho$. The set of transition points gives the numerical shadow boundary. Figure~\ref{fig:shadow_validation} compares the perturbative analytic shadow boundary with the full numerical ray-tracing result for a representative near-extremal black hole.
Their close agreement provides a nontrivial check of the finite-distance projection prescription and of the perturbative photon-region construction, since the numerical calculation uses the full metric without relying on the perturbative separability of the null Hamilton-Jacobi equation.

\begin{figure*}[t]
    \centering
    \includegraphics[width=0.3\textwidth]{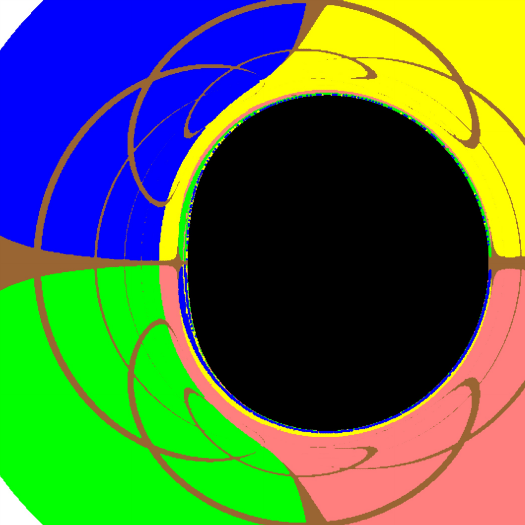}~~~
    \includegraphics[width=0.3\textwidth]{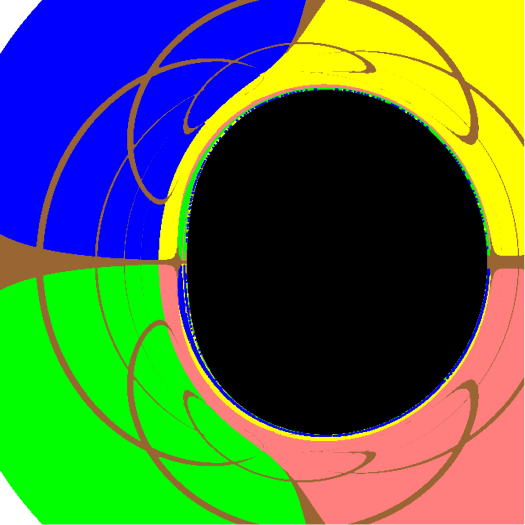}~~~
    \includegraphics[width=0.3\textwidth]{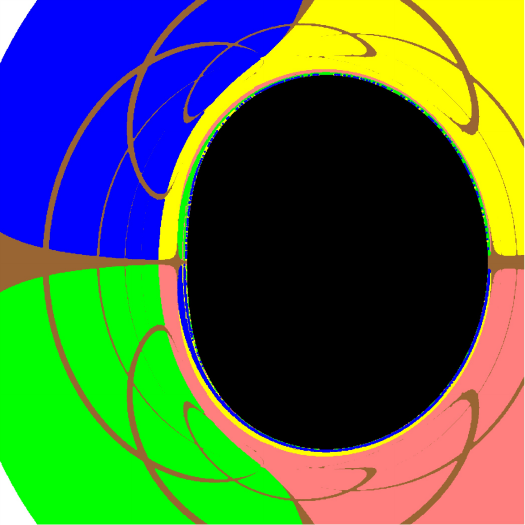}
    \caption{Ray-traced shadow images at fixed dimensionless spin $\chi=0.98$. From left to right, the panels correspond to $B=0$ (Kerr), $BM=0.005$, and $BM=0.01$, respectively. The horizontal and vertical axes are $\alpha/M$ and $\beta/M$, respectively. The observer radius and inclination angle are fixed at $r_o=50M$ and $\theta_o=90^\circ$ in all panels.}
    \label{fig:shadow_images}
\end{figure*}

Having established this agreement, we use direct ray tracing to generate representative shadow images for near-extremal black holes with the same dimensionless spin but different values of the spindle parameter, as shown in Fig.~\ref{fig:shadow_images}.
The $B=0$ case, corresponding to the Kerr black hole, serves as the benchmark, while the $B=0.005 $ and $B=0.01$ cases illustrate how the image is deformed as the spindle parameter is turned on and increased. Throughout this comparison, the dimensionless spin, observer radius, and inclination angle are kept fixed, so that the observed changes can be attributed to the spindle deformation. The figure shows that the spindle deformation affects not only the shadow boundary, which is controlled by the critical photon region, but also the surrounding lensing pattern, which reflects the propagation of nearby rays in the strong-field geometry.

\begin{figure*}[t]
    \centering
    \includegraphics[width=0.8\textwidth]{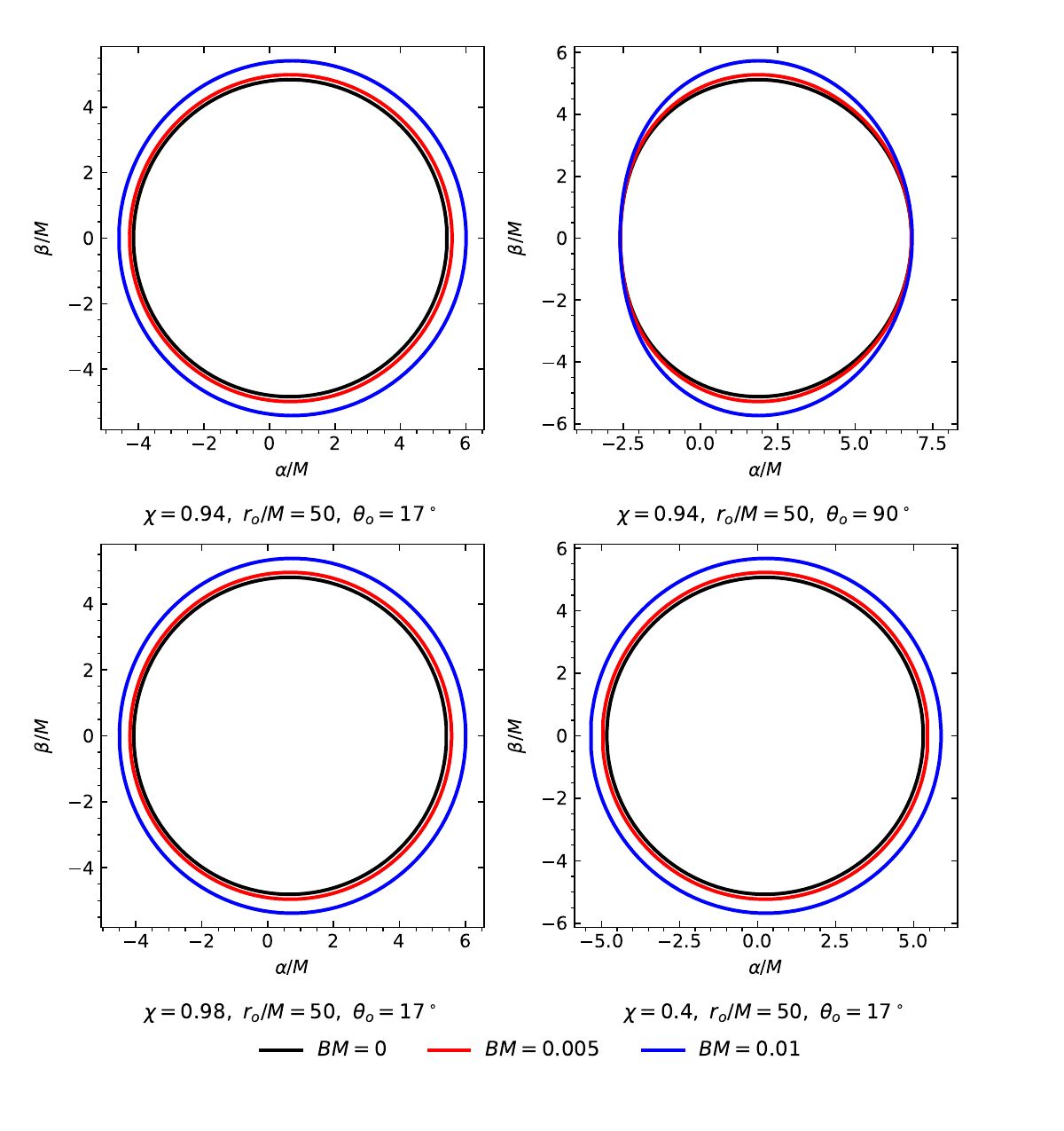}
    \caption{Analytic shadow boundaries obtained from the perturbative photon region at finite observer radius $r_o=50M$. Different panels correspond to different values of the dimensionless spin $\chi$ and the observer inclination angle $\theta_o$. In each panel, the Kerr shadow at $B=0$ is compared with ML shadows at nonzero $B$, while $\chi$, $r_o$, and $\theta_o$ are held fixed.}
    \label{fig:shadow_boundaries}
\end{figure*}

To explore the effect of the spindle deformation  on the shadow boundary more systematically, we plot in Fig.~\ref{fig:shadow_boundaries} the perturbative analytic shadow boundaries for different values of the dimensionless spin $\chi$ and the inclination angle $\theta_o$, with the observer radius fixed at $r_o=50M$. 
Figure~\ref{fig:shadow_boundaries} shows that changing the inclination angle affects both the apparent size and the shape of the shadow on the $(\alpha,\beta)$ screen. By contrast, varying the spin mainly increases the overall size of the shadow, while producing a comparatively weaker distortion of its shape. 

\begin{figure*}[t]
    \centering
    \includegraphics[width=0.9\textwidth]{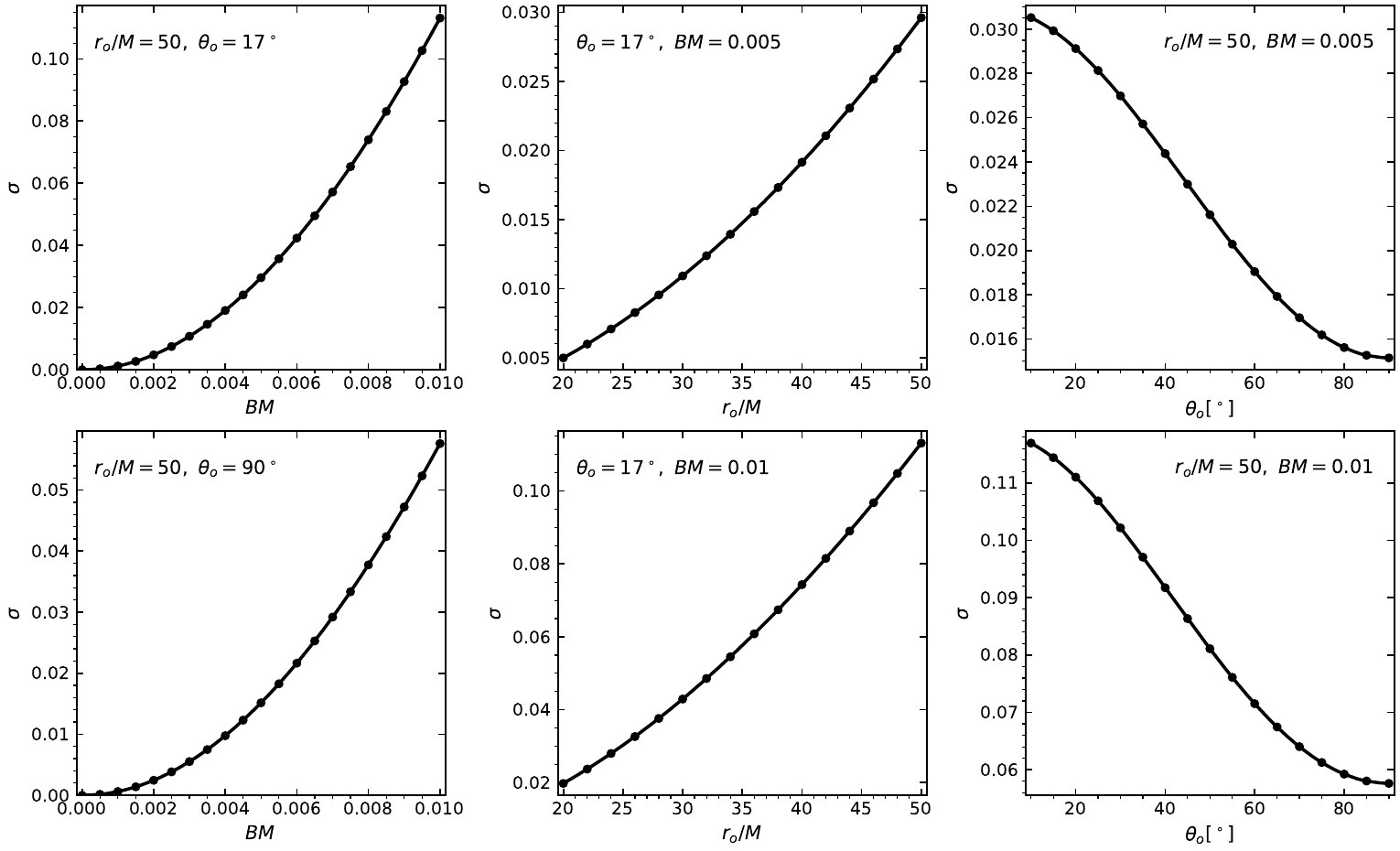}
    \caption{Dependence of the global Kerr-deviation parameter $\sigma=\bar\rho_{\rm ML}/\bar\rho_{\rm K}-1$ on the spindle parameter, observer radius, and inclination angle. In all cases, the ML shadow is compared with the Kerr shadow at the same dimensionless spin, observer radius, and inclination angle. The left, middle, and right columns show the dependence on $B$, $r_o$, and $\theta_o$, respectively.}
    \label{fig:shadow_deviation}
\end{figure*}

We finally quantify the deviation of the ML shadow from the Kerr benchmark. Let ${\cal C}_{\rm ML}$ and ${\cal C}_{\rm K}$ denote the ML and Kerr shadow boundaries on the same observer screen, computed with the same values of $\chi$, $r_o$, and $\theta_o$. For each closed shadow curve, we determine its centroid $(\alpha_c,\beta_c)$ from the enclosed area $A$ according to
\be
(\alpha_c,\beta_c)
=
\frac{1}{A}
\int_{\cal D}
(\alpha,\beta)
d\alpha d\beta \,,\quad
A
=
\int_{\cal D} d\alpha d\beta\,,
\label{areacentroid}
\ee
where ${\cal D}$ denotes the interior of the shadow. Measuring the boundary with respect to its centroid, we introduce
\be
\rho(\psi)
=
\sqrt{
\left[\alpha(\psi)-\alpha_c\right]^2
+
\left[\beta(\psi)-\beta_c\right]^2
}\,.
\label{rhofunction}
\ee
The mean shadow radius is then defined by
\be
\bar\rho
=
\frac{1}{2\pi}
\int_0^{2\pi}
\rho(\psi)d\psi\,.
\label{meanrho}
\ee
We characterize the global size deviation from Kerr by
\be
\sigma
=
\frac{\bar\rho_{\rm ML}}
{\bar\rho_{\rm K}}
-1\,.
\label{sigmadef}
\ee
This dimensionless quantity measures the fractional change in the mean shadow radius after the displacement of the shadow center has been removed. The results are shown in Fig.~\ref{fig:shadow_deviation}. 
In the small-deformation regime, the global size deviation increases with the spindle parameter and exhibits the expected absence of a term linear in $B$. Its dependence on the observer radius reflects the finite-distance character of the screen construction in the non-asymptotically flat ML geometry. The inclination dependence arises because changing $\theta_o$ modifies both the angular sector of the photon region accessible to the observer and the projection of the photon momenta onto the local screen. The quantity $\sigma$ therefore provides a finite-distance measure of the overall shadow-size deviation from the Kerr benchmark under the matching prescription adopted here.

\section{Conclusion} \label{conclusion}

Recently, a new exact Ricci-flat rotating black-hole solution was constructed, in which an additional parameter $B$ characterizes a spindle deformation of the Kerr geometry. In this work, we have studied how the spindle deformation affects geodesic motion and black-hole shadows in this new spacetime. Our main results are summarized as follows. First and foremost, unlike in Kerr, the Hamilton-Jacobi equation in the ML black hole is not exactly separable for either timelike or null geodesics. Remarkably, however, null geodesics become separable at the leading nontrivial order, ${\cal O}(B^2)$, whereas timelike geodesics remain nonseparable at the same order.

Second, for timelike motion, we studied equatorial circular geodesics and found that the spindle deformation shifts the ISCO outward and gives rise to an OSCO. As the deformation increases, the ISCO moves outward while the OSCO moves inward, so that the radial interval supporting stable circular motion gradually shrinks. The two marginally stable orbits eventually merge at a critical value of the deformation parameter, beyond which no stable circular orbit exists for the corresponding branch. 

Third, in the null sector, we used the perturbatively separated equations to analyze the photon region, equatorial photon orbits, and black-hole shadow, and validated the analytic results by direct ray tracing in the full spacetime. Similar to the timelike case, the spindle deformation shifts the equatorial photon orbits outward and increases the mean shadow radius relative to Kerr. We quantified these deviations for different values of the deformation parameter, spin, observer radius, and inclination angle. In the small-deformation regime, the numerical shadow boundary agrees well with the perturbative analytic result.

The loss of exact separability also points to a natural extension of the present analysis. Beyond the small-deformation regime, the Carter-like constant obtained at order ${\cal O}(B^2)$ is no longer expected to characterize the full null dynamics, while generic timelike motion is already nonseparable at this order. This suggests that the standard integrable structure of Kerr is lost and that generic geodesic motion may become nonintegrable. Since nonintegrable systems can exhibit chaotic behavior~\cite{Wang:2022kvg,Li:2026zxk,Dettmann:1994dj,Bombelli:1991eg,Sota:1995ms,Chen:2016tmr,Wang:2016wcj,Wang:2017qhh,Wang:2018eui}, the spindle-deformed geometry may also support chaotic particle motion. Chaotic motion of massive particles has already been studied in the static limit of the KBR spacetime~\cite{Li:2026zxk}. Whether analogous behavior occurs in the Ricci-flat $B$-deformed class of black holes deserves further investigation.



\begin{acknowledgments}
We are grateful to Minyong Guo, H. L\"u, and Liang Ma for useful discussions. 
H-D.L.~is  supported in part by National Natural Science Foundation of China Grants No.12447134 and Postdoctoral Innovation Project of Shandong Province SDCX-ZG-202503036.  
S.L. is supported in part by the National Natural Science Foundation of China (No. 12105098, No. 12481540179) and the Natural Science Foundation of Hunan Province (No. 2022JJ40264), and the innovative research group of Hunan Province under Grant No. 2024JJ1006, and by the Excellent Young Scholars Program of the Hunan Provincial Department of Education under Grant No. 25B0092. 
M. W. is supported in part by the National Natural Science Foundation of China under Grant No. 12105151, the Shandong Provincial Natural Science Foundation of China under Grant No. ZR2020QA080.

\end{acknowledgments}




\end{document}